\documentclass[preprint,12pt]{elsarticle}

\usepackage{amssymb}
\usepackage{amsmath}

\usepackage[dvipsnames,table,xcdraw]{xcolor}
\usepackage{tcolorbox}
\usepackage{multirow}
\usepackage{tabularx}
\usepackage{colortbl}
\usepackage{booktabs}
\usepackage{ragged2e}
\usepackage{url}
\newcolumntype{C}{>{\Centering\arraybackslash}X}

\journal{Computer Speech \& Language}

\begin{document}

\begin{frontmatter}


\title{LLM-based speaker diarization correction:\\A generalizable approach}

\author[label1,label2]{Georgios Efstathiadis}
 \ead{georgios.efstathiadis@brooklyn.health}
\author[label1,label3]{Vijay Yadav}
 \ead{vijay.yadav@brooklyn.health}
\author[label1]{Anzar Abbas}
 \ead{anzar@brooklyn.health}

\affiliation[label1]{organization={Brooklyn Health},
            addressline={11201 USA}, 
            city={Brooklyn},
            state={NY},
            country={USA}}

\affiliation[label2]{organization={Department of Biostatistics, Harvard T. H. Chan School of Public Health},
            addressline={02115 USA}, 
            city={Boston},
            state={MA},
            country={USA}}

\affiliation[label3]{organization={School of Psychology, University of New South Wales},
            addressline={2052 Australia}, 
            city={Sydney},
            state={NSW},
            country={Australia}}

\begin{abstract}
Speaker diarization is necessary for interpreting conversations transcribed using automated speech recognition (ASR) tools. Despite significant developments in diarization methods, diarization accuracy remains an issue. Here, we investigate the use of large language models (LLMs) for diarization correction as a post-processing step. LLMs were fine-tuned using the Fisher corpus, a large dataset of transcribed conversations. The ability of the models to improve diarization accuracy in a holdout dataset from the Fisher corpus as well as an independent dataset was measured. We report that fine-tuned LLMs can markedly improve diarization accuracy. However, model performance is constrained to transcripts produced using the same ASR tool as the transcripts used for fine-tuning, limiting generalizability. To address this constraint, an ensemble model was developed by combining weights from three separate models, each fine-tuned using transcripts from a different ASR tool. The ensemble model demonstrated better overall performance than each of the ASR-specific models, suggesting that a generalizable and ASR-agnostic approach may be achievable. We have made the weights of these models publicly available on HuggingFace at \url{https://huggingface.co/bklynhlth}.
\end{abstract}




\end{frontmatter}



\section{Introduction}
Diarization refers to the identification of unique speakers in a conversation \cite{anguera-2012, tranter-2003}. It is often a component of automated speech recognition (ASR) tools \cite{aws_transcribe, azure_transcribe}. It is necessary in various contexts, such as medical transcriptions \cite{finley-2018, mirheidari-2017}, where separating patient and clinician speech is necessary for interpretation. Diarization accuracy is impacted by several factors such as audio quality, environmental noise, variability in speaker behavior, and overlapping speech \cite{khazaleh-2024, godin-2015, charlet-2013}. Here, we present a method that uses a fine-tuned large language model (LLM) to improve diarization accuracy in conversational transcripts.

Several speaker diarization methods exist \cite{park-2022b, serafini-2023}. Traditionally, clustering‐based approaches have been widely used. One such method is the x-vector approach \cite{snyder-2018}, which extracts speaker embeddings, or x-vectors, from fixed-length segments of audio using a time-delay neural network. The embeddings are clustered to assign speaker labels. Building on these foundations, earlier neural network–based methods further improved diarization performance. A popular open source method is Pyannote \cite{bredin-2020}, which uses pre-trained neural networks to assign speaker labels through analysis of acoustic data. It processes audio into overlapping segments, extracting features such as mel-frequency cepstral coefficients to detect changes in speakers.

More recently, the field has shifted towards end-to-end diarization models \cite{fujita-2019b, zhang-2019, kanda-2022, zheng-2022, meng-2023, cornell-2024} which integrate speaker recognition with transcription, allowing for simultaneous transcription and diarization. These models map audio inputs directly to speaker-attributed transcriptions. Fujita et al. \cite{fujita-2019b} introduced a self-attention mechanism to enhance the End-to-End Neural Diarization (EEND) model \cite{fujita-2019a}, showing improved performance over traditional BLSTM-based methods by directly optimizing the diarization error rate (DER), which significantly reduces errors in overlapping speech scenarios. Zhang et al. \cite{zhang-2019} proposed a fully supervised diarization system named unbounded interleaved-state recurrent neural network (UIS-RNN), which replaces traditional unsupervised clustering modules with an online generative process, leading to significant improvements in diarization error rates, particularly when trained with high-quality, time-stamped data. Kanda et al. proposed the transcribe-to-diarize model \cite{kanda-2022}, which applies a neural network that processes audio to provide both the transcribed text and the speaker labels in one pass with the ability to estimate the start and end times of each word by incorporating a minimal set of learnable parameters into the model's internal state.

Diarization is often a component of popular ASR tools such as those offered by Amazon Web Services (AWS), Azure, or Google Cloud Platform (GCP). These ASRs tools have various distinguishing features such as transcription accuracy, language support, inference speed, and – relevant to our manuscript – the diarization method employed. While some popular ASR tools such as WhipserX are open source \cite{bain-2023}, others do not disclose details on diarization methods. In a given use case, depending on the ASR tool used, differences in diarization accuracy may influence downstream findings and interpretation \cite{ferraro-2023, xu-2021}. Where addressing lack of visibility into underlying methods may not be practical, methods to improve and consequently converge diarization accuracy in transcripts from different ASR tools would allow for ASR-agnostic applications.

\begin{figure}[!t]
\centering
\includegraphics[width=\textwidth]{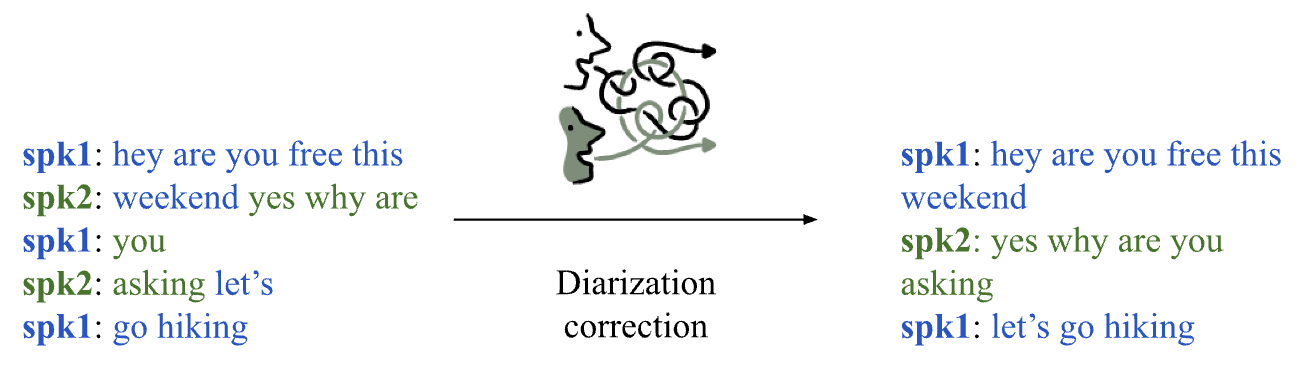}
\caption{Accurate speaker diarization is necessary for interpretation of important conversations.}
\label{fig1}
\vspace{-1em}
\end{figure}

Several methods have been explored to improve speaker diarization accuracy. Han et al. 2024 \cite{han-2024} proposes a network architecture to improve speaker diarization from an initial diarization attempt. Their model incorporates two parallel encoder networks: one processes the initial diarization results; the other analyzes the log-mel frequency spectrum of the audio. These are integrated by a decoder that merges the output embedding from both encoders. Paturi et al. 2023 \cite{paturi-2023} uses a similar approach. They fine tune an encoder network that utilizes the embedding. From a pre-trained language model combined with the original speaker ID embeddings to recalibrate the speaker IDs. Duke et al. 2022 \cite{duke-2022} employs a theoretical approach by modeling the probability of an utterance being assigned to a speaker using Bayesian inference. The probability is calculated based on the semantic distance between an utterance and its preceding utterances.

Other methods have utilized LLMs to improve speaker diarization. Wang et al. 2024 \cite{wang-2024} established the fine-tuning framework employed in this manuscript. They fine-tune PaLM 2-S \cite{anil-2023} to correct speaker diarization mistakes from GCP’s Universal Speech Model \cite{zhang-2023}, which uses Turn-to-Diarize \cite{xia-2022} for speaker diarization. Their manuscript shows that the approach can notably enhance speaker diarization accuracy. Park et al. 2024 \cite{park-2024} fine-tuned an LLM to correct speaker diarization of an acoustic-only ASR tool, a version of the Multi-Scale Diarization Decoder model \cite{park-2022}. They combine LLM predictions with acoustic information to build a beam search decoding approach that showed improvement in speaker labeling. Adedji et al. 2024 \cite{adedeji-2024} used pre-trained LLMs in a multi-step approach to improve speaker diarization in medical transcripts. By using a separate component to transcribe the text and an LLM to label speakers, they showed that using chain-of-thought speaker prompts can diarization ASR-transcribed text and match or surpass the performance of integrated ASR tools that conduct both transcription and diarization. 

Here, we fine-tune LLMs to improve diarization accuracy in ASR transcripts as a post-processing step, visualized in Figure \ref{fig1}. To fine-tune our models, we used the English Fisher corpus \cite{cieri-2004}, a large audio dataset containing 1,960 hours of phone conversations, and to test them we used a held out set of the Fisher corpus and the PriMock57 dataset \cite{korfiatis-2022} a smaller audio dataset containing mock primary care consultations. We consider the effects of using transcripts from different ASRs, tuning separate models for each AWS, Azure, and WhisperX transcript. Wang et al. (2024) demonstrated notable improvements in speaker diarization accuracy through fine-tuning. Our work investigates the generalizability of this method across different ASR systems. Specifically, we identify a key limitation in previous approaches: the variability in performance when applied to different ASRs. We find that LLM-based diarization correction models are most effective when correcting transcripts produced using the same ASR as the transcripts they were fine-tuned with, limiting their generalizability. To address this constraint, we build an ensemble model that combines the weights from each of the individual models, and demonstrate that such a model has the potential to serve as a reliable and ASR-agnostic diarization correction tool.

\begin{figure}[!t]
\centering
\includegraphics[width=0.4\textwidth]{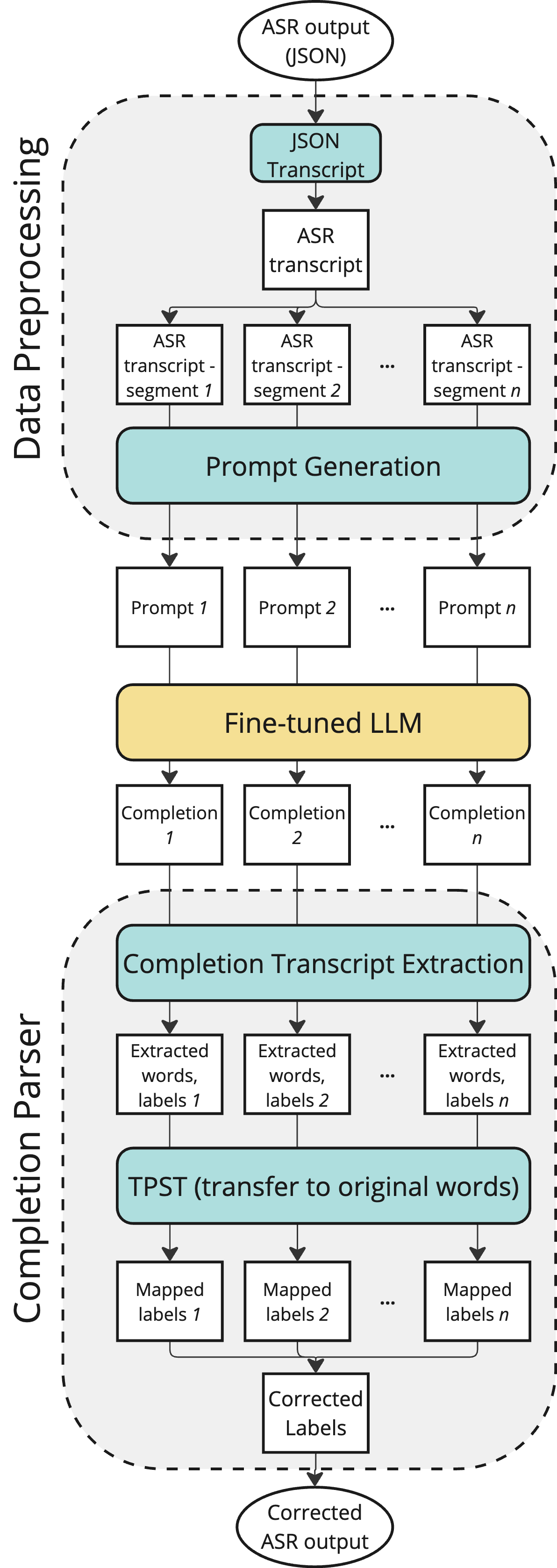}
\caption{Overview of pipeline used to evaluate the LLM.}
\label{fig3}
\vspace{-1em}
\end{figure}

This model has been published on HuggingFace and is available at \url{https://huggingface.co/bklynhlth/WillisDiarize-v1}. We additionally released a function using the methods described here for a friendlier user interface on applying diarization correction in OpenWillis \cite{worthington-2024}, a Python library for digital measurement of health, that can be found at \url{https://github.com/bklynhlth/openwillis}. An overview of the system used in a real-world application can be seen in Figure \ref{fig3}.

\section{Methods}

Code for all data, training and evaluation methods can be found at \url{https://github.com/GeorgeEfstathiadis/LLM-Diarize-ASR-Agnostic}.

\subsection{Data}

The English Fisher corpus, available through the Linguistic Data Consortium, was used \cite{cieri-2004}. The dataset contains 11,699 audio recordings of telephone conversations between two participants. The recordings total 1,960 hours in length, averaging 10 minutes each. 11,971 participants (57\% female, average age 36 years) were involved, each participating in 1.95 calls on average. Recordings in the Fisher corpus have corresponding transcripts with sentence-level timestamps. The transcribed words and speaker labels are referred to here as the \textbf{reference transcripts}. Methods for production of these transcripts can be found in Cieri et al. 2004 \cite{cieri-2004}. Data was split into a training and testing set using the same split as in past work \cite{wang-2024, wang-2022, zhao-2023A, zhao-2023B}: The training set had 11,527 recordings from 11,631 participants; the testing set had 172 recordings from 340 participants.

In addition we tested the model on an independent dataset. We used PriMock57 for evaluation  \cite{korfiatis-2022}, a dataset of 57 mock primary consultations between a clinician and a mock patient. The recordings total approximately 9 hours in length, averaging 9 minutes each. There were 7 clinicians and 57 actors portraying the patient involved. Same as in the Fisher corpus, the PriMock57 dataset had corresponding transcripts with sentence-level timestamps. Methods for how these mock consultations were conducted can be found in Korfiatis et al. 2022 \cite{korfiatis-2022}.

\subsection{Transcription}

All recordings were transcribed into text using three ASR tools: AWS’s Transcribe \cite{aws_transcribe}; Azure’s Speech to Text \cite{azure_transcribe}; and WhisperX \cite{bain-2023}. Each ASR conducts both transcription and diarization. WhisperX is a combination of Whisper, OpenAI’s open-source transcription model \cite{radford-2022}, and Pyannote’s open-source diarization model \cite{bredin-2020}.

When running transcriptions through both AWS and Azure, language was prespecified as English, the maximum number of speakers was set to 2, and the option of speaker diarization was set to true. Version 3.2 of Azure’s Speech to Text API was used. When running transcriptions through WhisperX, the Whisper large-v2 model was used for transcription, Pyannote version 3.1.1 was used for diarization, the language was prespecified as English, and the maximum number of speakers were set to 2. All transcriptions produced a JSON file for each recording, containing the transcribed text with speaker labels.

\subsection{Data Preprocessing}

Given that each JSON file had ASR-specific formatting, we extracted transcribed words and speaker labels into a common format, consistent with what was used in Wang et al. 2024 \cite{wang-2024}. These transcripts are referred to here as the \textbf{ASR transcripts}. The ASR transcripts were further standardized by removing punctuation and transforming all text to lowercase. 11\% of reference transcripts were missing transcripts for part of the recording, ranging from 30 seconds up to 8 minutes and 27 seconds. For these, we trimmed the ASR transcripts to correspond with their reference transcripts.

While all input audio recordings contained conversation of 2 speakers, in some cases ASRs may only detect 1 speaker. This can be due to poor audio quality, noise in the recording, volume of the speakers or other factors. Any ASR transcripts with only one or more than two detected speakers were removed from the dataset. Some ASR transcripts, particularly those from WhisperX, had continuously repeating word sequences \cite{koenecke-2024}. For example, words such as ‘and’ or phrases such as ‘that’s right’ would repeat consecutively such that the sentence would lose meaning. To minimize the effect of repeating sequences on model fine-tuning and testing, transcripts that contained more than 10 repetitions of a sequence were removed. 

\begin{figure*}[!t]
\centering
\includegraphics[width=\textwidth]{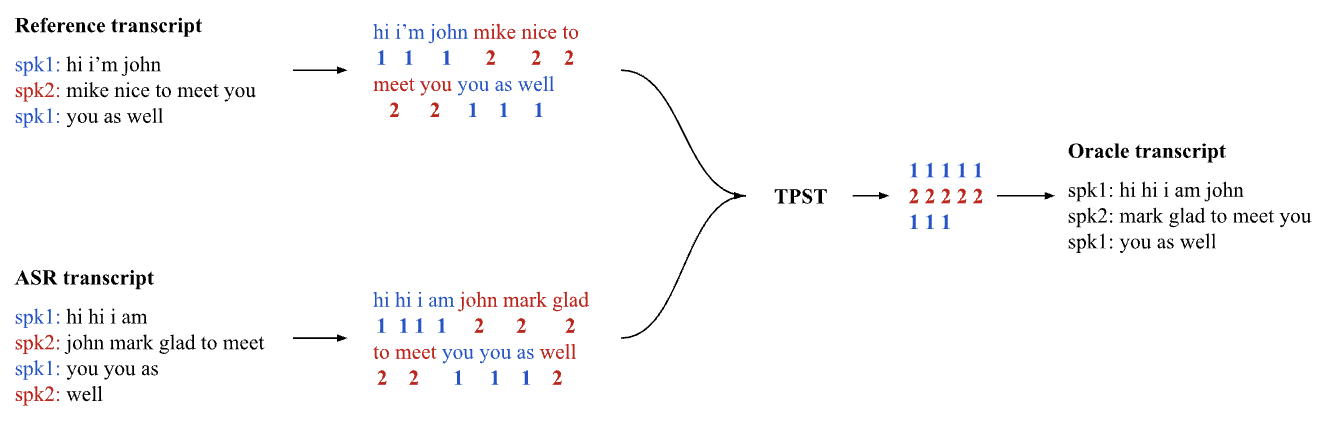}
\caption{Creation of oracle transcripts using the TPST algorithm. Words and speaker labels are extracted from each transcript. The algorithm aligns word sequences, such that the resulting speaker labels from the reference transcript match the text of the ASR transcript. This corrects speaker labeling in the ASR transcript without changing the underlying transcription.}
\label{fig2}
\end{figure*}

To create the training data, a transcript-preserving speaker transfer algorithm (TPST) \cite{wang-2024} was used to transfer speaker labels from the reference transcripts to the ASR transcripts shown in Figure \ref{fig2}. The TPST algorithm accepts speaker-labeled source text and speaker-labeled target text. It aligns the two so that the source text speaker labels match the target text. The resulting transcripts are referred to here as the \textbf{oracle transcripts}. Oracle transcripts have the same text as the ASR transcripts but the speaker labels of the reference transcripts. They allow for comparison of diarization accuracy between an ASR transcripts and the ground-truth speaker labeling in the reference transcripts by removing effects for differences in the transcription itself.

\subsection{Model Tuning}

\subsubsection{Model Selection}

We used Mistral AI’s Mistral 7b Instruct v0.2 as the base model for all fine-tuned models in this manuscript \cite{jiang-2023}. Compared to similar open-source large language models, Mistral 7b was chosen because of its size and relative baseline accuracy in diarization correction. The decision-making process behind model selection is presented in \ref{app1}.

\subsubsection{ASR-specific models}

Given each ASR tool has a noticeably different transcription style, in terms of both words utilized and also diarization errors made, we hypothesized that a model fine-tuned on one ASR would perform best at correcting diarization for transcripts produced by the same ASR compared to transcripts produced by a different ASR. For this reason, three separate ASR-specific models were fine-tuned using AWS, Azure, and WhisperX transcripts.

Fine-tuning involved using the ASR transcripts in the prompt and the oracle transcripts in the completion, using the same method as Wang et al. 2024 \cite{wang-2024}. The fine-tuning prompt and completion template is shown below; the exact template can be found in \url{https://github.com/GeorgeEfstathiadis/LLM-Diarize-ASR-Agnostic}.

\begin{tcolorbox}[colback=gray!10, colframe=gray!40, title=]
\textbf{Prompt:}

\vspace{1em}

In the speaker diarization transcript below, some words are potentially misplaced. Please correct those words and move them to the right speaker. Directly show the corrected transcript without explaining what changes were made or why you made those changes:

\vspace{1em}

\texttt{[ASR transcript]}

\vspace{1em}

\textbf{Completion:}

\vspace{1em}

\texttt{[Oracle transcript]}
\end{tcolorbox}

Though the context length of Mistral 7b is 32 thousand tokens, we segmented transcripts in the training set to achieve prompt-completion pairs totaling 8,192 tokens. This was because smaller segments ensure efficient memory usage and align better with practical use-case scenarios.

We fine-tuned each of the three ASR-specific models using Amazon SageMaker on an ml.g5.8xlarge instance (1 NVIDIA A10G Tensor Core GPU with 24 GiB of memory) for 2 epochs, with a batch size of 6 and 5 gradient accumulation steps. We used Quantized Low-Ranked Adaption of Language Models \cite{dettmers-2023} and Flash attention \cite{dao-2022} for efficient fine-tuning, balancing computational resources with model performance improvement. The three resulting fine-tuned models are referred to here as the \textbf{AWS model}, \textbf{Azure model}, and \textbf{WhisperX model}, or collectively as the \textbf{ASR-specific models}.

\subsubsection{Ensemble model}

To develop an ASR-agnostic ensemble model, TIES-Merging \cite{yadav-2023} from merge-kit \cite{goddard-2024} was used to combine parameters from each ASR-specific model. This computes a weighted average of each parameter, such that only the most significant changes of each parameter are used, with the rest set back to the base model value, and conflicts between different models are solved by using the most significant change across different versions. For hyperparameters, we set all models to approximately equal weight (AWS: 0.34, Azure: 0.33, WhisperX: 0.33) and the density for each model to 0.8.

\subsection{Completion parser}

A completion parser was developed to post-process the model output. It extracts only the relevant part from the model output, removing pre- and post-fixes if present, extracts speaker labels from the model output, and transfers those labels back to input text. Since LLMs can hallucinate, it is common that the LLM may alter original wording in the transcript in addition to speaker labeling. To mitigate this, the TPST algorithm––the same algorithm as seen in creating the oracle transcripts during training––was used to ensure speaker labels outputted by the model matched the original wording. This is done by transferring the completion speaker labels to the input ASR transcription wording, thus mitigating word changes made by the LLM in the completion wording. This completion parser was integrated as part of the model’s methods to ensure no word substitutions or modifications were made through use of the model.

\subsection{Model evaluation}

\subsubsection{Measurement of accuracy}

Diarization accuracy was measured using two established metrics: delta concatenated minimum-permutation word error rate (deltaCP) \cite{watanabe-2020} and delta speaker-attributed word error rate (deltaSA) \cite{cornell-2023}. For each, word error rate (WER) first needed to be calculated. WER was measured as the percentage of transcription errors divided by the total number of words. When measuring accuracy in this manuscript, deltaCP and deltaSA were calculated with the reference transcripts serving as ground truth. Together, deltaCP and deltaSA quantify how much additional error is introduced by speaker labeling, allowing us to differentiate between transcription errors and those stemming from misattributed speakers.

To calculate deltaCP, concatenated minimum-permutation word error rate (cpWER) was first calculated. To do this, each speaker’s words were separated into their own transcripts. For each possible permutation of the speaker’s words, the WER was calculated against the reference speaker transcript. The smallest observed WER for each of the two speakers was averaged. cpWER is considered the minimum possible WER among all permutations of the two speakers. deltaCP is simply cpWER subtracted by the original WER (i.e. deltaCP = cpWER - WER). The difference allows us to analyze errors introduced by speaker labeling, independent of the original WER.

To calculate deltaSA, speaker-attributed word error rate (SA-WER) was first calculated. To do this, each speaker’s words were separated into their own transcripts. WER was calculated against the reference speaker-transcripts. The WER of each of the two speakers was averaged. deltaSA is simply SA-WER subtracted by the WER (i.e. deltaSA = SA-WER - WER).

\subsubsection{Assessing performance}

Diarization correction was performed on the ASR transcripts in the testing set. This was done separately for transcripts from each of the three ASRs. The resulting transcripts are referred to here as the \textbf{corrected transcripts}. Before diarization correction, transcripts were segmented into smaller chunks to meet the 4,096 token threshold (8,192 tokens for combined prompt-completion).

Model performance was assessed by calculating deltaCP and deltaSA before and after diarization correction. Seven different models for diarization correction were used:

\begin{enumerate}
\item{Mistral 7b, with no fine-tuning}
\item{Mistral 8x7b, with no fine-tuning \cite{jiang-2024}}
\item{DiarizationLM 8b Fisher v2 \cite{wang-2024}}
\item{Mistral 7b fine-tuned on AWS transcripts (AWS model)}
\item{Mistral 7b fine-tuned on Azure transcripts (Azure model)}
\item{Mistral 7b fine-tuned on WhisperX transcripts (WhisperX model)}
\item{The ensemble model built using the three ASR-specific models}
\end{enumerate}

DiarizationLM 8b Fisher v2 is the best performing model from the models that are made openly available, proposed by Wang et al. 2024 \cite{wang-2024}. The framework they employed is the one we based our training strategy on for our ASR-specific models. The base model used was Llama3 with 8b parameters which is similar in size with the base model we used which had 7b parameters.

When assessing performance of the fine-tuned models, the same prompt from model training was used. For the zero-shot performance assessment i.e. for Mistral 7b and Mistral 8x7b with no-finetuning, the following prompt was used:

\begin{tcolorbox}[colback=gray!10, colframe=gray!40, title=]
\textbf{Prompt:} \\

\texttt{<s>[INST]} \\
In the speaker diarization transcript below, some words are potentially misplaced. Please correct those words and move them to the right speaker. Directly show the corrected transcript without explaining what changes were made or why you made those changes: \\
\\
\texttt{[ASR transcript] [/INST]} \\
\\
Here is the corrected transcript with the words moved to the right speaker:
\end{tcolorbox}

In addition, when evaluating DiarizationLM 8b Fisher v2 which was proposed and open-sourced in Wang et al. 2024 \cite{wang-2024}, the following prompt template was used (as shown at https://huggingface.co/google/DiarizationLM-8b-Fisher-v2):

\begin{tcolorbox}[colback=gray!10, colframe=gray!40, title=]
\textbf{Prompt:} \\

\texttt{[ASR transcript]} $\rightarrow$
\end{tcolorbox}

\section{Results}

\subsection{Transcripts used}

After removing transcripts with only one or more than two detected speakers, we were left with 11,519 transcripts in the AWS training set (8 removed), 11,525 transcripts in the Azure training set (2 removed), and 11,464 transcripts in the WhisperX training set (63 removed).

After splitting the transcripts into smaller segments to fit the 4,096 token context window, the training set included 52,293 AWS transcripts, 58,192 Azure transcripts, and 53,997 WhisperX transcripts, each with a corresponding oracle transcript.

Then, the transcripts with repeating word sequences were removed. This was done after the segmentation to avoid removing more data than necessary. The final training set included 52,287 AWS transcripts (6 removed), 58,184 Azure transcripts (8 removed), and 52,982 WhisperX transcripts (1,015 removed).

The same steps were carried out on the Fisher testing set and the PriMock57 dataset. After preprocessing, for the Fisher testing set we were left with 172 AWS transcripts, 172 Azure transcripts (both AWS and Azure did not have any transcripts removed), and 162 WhisperX transcripts (1 removed due to a single speaker and 9 removed due to repeating word sequences). While for the PriMock57 dataset we were left with 57 AWS transcripts (0 removed), 57 Azure transcripts (0 removed) and 43 WhisperX transcripts (14 removed due to single or more than 2 speakers).

\subsection{Word error rate}

We looked at WER across the three ASRs. The WER across ASRs is shown in Table \ref{tab:base_measures}. The PriMock57 dataset had considerably better audio quality, given the mock sessions were carefully conducted in an isolated environment, which explains the improved transcription WER compared to the Fisher test set. Azure’s Speech to Text model performed best in terms of transcription accuracy in the Fisher test set, while AWS’s Transcribe performed the best in the PriMock57 dataset. WhisperX performed the worst in both datasets.

\subsection{Baseline diarization accuracy}

Diarization accuracy was compared between ASR transcripts and corresponding reference transcripts without any diarization correction. The baseline diarization accuracy across ASRs is shown in Table \ref{tab:base_measures}. In the Fisher dataset, AWS performed best, while Azure was next and WhisperX performed worst. On the PriMock57 dataset Azure performed the best and AWS was next. WhisperX performed worst on the PriMock57 dataset as well. The difference in diarization performance between AWS’s Transcribe and Azure’s Speech to Text is indicative of the fact that different ASRs may work better for different use cases.

\begin{table*}[ht]
\centering
\caption{Baseline transcription and diarization accuracy measures across ASRs.}
\label{tab:base_measures}
\setlength{\tabcolsep}{4pt}
\tiny
\begin{tabularx}{\textwidth}{l*{9}{C}}
\toprule
\multirow{2}{*}{} & \multicolumn{3}{c}{\textbf{AWS transcripts}} & \multicolumn{3}{c}{\textbf{Azure transcripts}} & \multicolumn{3}{c}{\textbf{WhisperX transcripts}} \\ \cmidrule(lr){2-10}
                  & deltaCP & deltaSA & WER & deltaCP & deltaSA & WER & deltaCP & deltaSA & WER \\ \midrule
\textbf{Fisher test set} & 0.93 & 2.5 & 22.04 & 1.91 & 3.06 & 16.99 & 4.46 & 5.77 & 22.39 \\
\textbf{PriMock57 dataset}  & 2.18 & 3.56 & 12.43 & 0.82 & 1.97 & 14.28 & 4.96 & 6.56 & 18.69
 \\ \bottomrule
\end{tabularx}
\end{table*}

\subsection{Zero-shot model performance}

The performance of the Mistral 7b and Mistral 7x8b models in improving diarization accuracy was measured against baseline, i.e, transcripts with no diarization correction. This was done using the Fisher test dataset only. The results are shown in Table \ref{tab:mistral}.

\begin{table*}[ht]
\centering
\caption{Zero-shot evaluation of base models with no fine-tuning on correction of diarization errors in the Fisher test set.}
\label{tab:mistral}
\setlength{\tabcolsep}{4pt}
\tiny
\begin{tabularx}{\textwidth}{l*{12}{C}}
\toprule
\multirow{3}{*}{} & \multicolumn{4}{c}{\textbf{AWS transcripts}} & \multicolumn{4}{c}{\textbf{Azure transcripts}} & \multicolumn{4}{c}{\textbf{WhisperX transcripts}} \\ \cmidrule(lr){2-13}
                  & \multicolumn{2}{c}{deltaCP} & \multicolumn{2}{c}{deltaSA} & \multicolumn{2}{c}{deltaCP} & \multicolumn{2}{c}{deltaSA} & \multicolumn{2}{c}{deltaCP} & \multicolumn{2}{c}{deltaSA} \\ \cmidrule(lr){2-13}
                  & Value & $\Delta$ & Value & $\Delta$ & Value & $\Delta$ & Value & $\Delta$ & Value & $\Delta$ & Value & $\Delta$ \\ \midrule
\textbf{Baseline} & 0.93 &  & 2.5 &  & 1.91 &  & 3.06 &  & 4.46 &  & 5.77 &  \\ \midrule
\textbf{Mistral 7b}  & 16.53 & \textcolor{Red}{+1677\%} & 19.28 & \textcolor{Red}{+671\%} & 16.85 & \textcolor{Red}{+782\%} & 19.43 & \textcolor{Red}{+535\%} & 21.26 & \textcolor{Red}{+377\%} & 24.09 & \textcolor{Red}{+318\%} \\
\textbf{Mistral 7x8b} & 12.01 & \textcolor{Red}{+1191\%} & 14.44 & \textcolor{Red}{+478\%} & 12.88 & \textcolor{Red}{+574\%} & 15 & \textcolor{Red}{+390\%} & 17.5 & \textcolor{Red}{+292\%} & 19.94 & \textcolor{Red}{+246\%} \\ \bottomrule
\end{tabularx}
\end{table*}

The base models, with no fine-tuning, i.e., the zero-shot models, instead of improving diarization accuracy, significantly reduced diarization accuracy compared to baseline.

\subsection{Fine-tuned model performance}

The performance of the ASR-specific models and the ensemble model in improving diarization accuracy was measured against baseline in the Fisher test set. We also evaluated DiarizationLM 8b Fisher v2 \cite{wang-2024}. The results are shown in Table \ref{tab:evaluation}.

\begin{table*}[ht]
\centering
\caption{Evaluation of fine-tuned models on correction of diarization errors in the Fisher test set. ASR-specific models performed best on transcripts produced using the same ASR as the transcripts used for fine-tuning. The ensemble model performed best across the board, regardless of the ASR used to produce the transcript.}
\label{tab:evaluation}
\renewcommand{\arraystretch}{1.2}
\setlength{\tabcolsep}{4pt}
\tiny
\begin{tabularx}{\textwidth}{l*{12}{C}}
\toprule
\multirow{3}{*}{} & \multicolumn{4}{c}{\textbf{AWS transcripts}} & \multicolumn{4}{c}{\textbf{Azure transcripts}} & \multicolumn{4}{c}{\textbf{WhisperX transcripts}} \\ \cmidrule(lr){2-13}
                  & \multicolumn{2}{c}{deltaCP} & \multicolumn{2}{c}{deltaSA} & \multicolumn{2}{c}{deltaCP} & \multicolumn{2}{c}{deltaSA} & \multicolumn{2}{c}{deltaCP} & \multicolumn{2}{c}{deltaSA} \\ \cmidrule(lr){2-13}
                  & Value & $\Delta$ & Value & $\Delta$ & Value & $\Delta$ & Value & $\Delta$ & Value & $\Delta$ & Value & $\Delta$ \\ \midrule
\textbf{Baseline} & 0.93 &  & 2.5 &  & 1.91 &  & 3.06 &  & 4.46 &  & 5.77 &  \\ \midrule
\textbf{DiarizationLM} & 3.94 & \textcolor{Red}{+324\%} & 5.71 & \textcolor{Red}{+128\%} & 5.97 & \textcolor{Red}{+213\%} & 7.3 & \textcolor{Red}{+139\%} & 6.05 & \textcolor{Red}{+36\%} & 7.46 & \textcolor{Red}{+29\%} \\ \midrule
\textbf{AWS model}      & \cellcolor{gray!20}\textbf{0.5} & \cellcolor{gray!20}\textcolor{Green}{\textbf{-46\%}} & \cellcolor{gray!20}\textbf{2.05} & \cellcolor{gray!20}\textcolor{Green}{\textbf{-18\%}} & 1.3 & \textcolor{Green}{-32\%} & 2.41 & \textcolor{Green}{-22\%} & 4.11 & \textcolor{Green}{-8\%} & 5.39 & \textcolor{Green}{-7\%} \\
\textbf{Azure model}    & 1.04 & \textcolor{Red}{+12\%} & 2.6 & \textcolor{Red}{+4\%} & \cellcolor{gray!20}0.87 & \cellcolor{gray!20}\textcolor{Green}{-54\%} & \cellcolor{gray!20}1.92 & \cellcolor{gray!20}\textcolor{Green}{-37\%} & 3.89 & \textcolor{Green}{-13\%} & 5.12 & \textcolor{Green}{-11\%} \\
\textbf{WhisperX model} & 1.71 & \textcolor{Red}{+84\%} & 3.44 & \textcolor{Red}{+38\%} & 1.46 & \textcolor{Green}{-24\%} & 2.57 & \textcolor{Green}{-16\%} & \cellcolor{gray!20}3.37 & \cellcolor{gray!20}\textcolor{Green}{-24\%} & \cellcolor{gray!20}4.61 & \cellcolor{gray!20}\textcolor{Green}{-20\%} \\ \midrule
\textbf{Ensemble model} & 0.63 & \textcolor{Green}{-32\%} & 2.15 & \textcolor{Green}{-14\%} & \textbf{0.82} & \textcolor{Green}{\textbf{-57\%}} & \textbf{1.88} & \textcolor{Green}{\textbf{-39\%}} & \textbf{3.15} & \textbf{\textcolor{Green}{-29\%}} & \textbf{4.31} & \textcolor{Green}{\textbf{-25\%}} \\ \bottomrule
\end{tabularx}
\end{table*}

DiarizationLM 8b Fisher v2 is unable to improve diarization performance in unseen ASRs. Fine-tuned models markedly improved diarization accuracy compared to baseline in most cases. 

As hypothesized, each of the ASR-specific models improved diarization accuracy most significantly on transcripts that were produced using the same ASR tool as the transcripts in the training set. The ASR-specific models did not perform as well on transcripts derived from a different ASR, with some worsening diarization accuracy.

The ensemble model achieved the best overall performance. In the AWS transcripts, it had marginally worse performance compared to the AWS model. However, in the Azure and WhisperX transcripts, it performed better than even the ASR-specific models for Azure and WhisperX. Ablation studies on ensembles of two ASR-specific models indicate that while partial improvements can be achieved by combining subsets of ASR-specific models, the full three-model ensemble provides the best generalization across different ASR outputs (see \ref{app3} for details).

\subsection{Performance on an independent dataset}

We also evaluated the fine-tuned models on the PriMock57 dataset. The results are shown in Table \ref{tab:primock57}.

\begin{table*}[ht]
\centering
\caption{Evaluation of fine-tuned models on correction of diarization errors in the PriMock57 dataset. ASR-specific models performed best on transcripts produced using the same ASR as the transcripts used for fine-tuning in most cases. The ensemble model performed best across the board, regardless of the ASR used to produce the transcript.}
\label{tab:primock57}
\renewcommand{\arraystretch}{1.2}
\setlength{\tabcolsep}{4pt}
\tiny
\begin{tabularx}{\textwidth}{l*{12}{C}}
\toprule
\multirow{3}{*}{} & \multicolumn{4}{c}{\textbf{AWS transcripts}} & \multicolumn{4}{c}{\textbf{Azure transcripts}} & \multicolumn{4}{c}{\textbf{WhisperX transcripts}} \\ \cmidrule(lr){2-13}
                  & \multicolumn{2}{c}{deltaCP} & \multicolumn{2}{c}{deltaSA} & \multicolumn{2}{c}{deltaCP} & \multicolumn{2}{c}{deltaSA} & \multicolumn{2}{c}{deltaCP} & \multicolumn{2}{c}{deltaSA} \\ \cmidrule(lr){2-13}
                  & Value & $\Delta$ & Value & $\Delta$ & Value & $\Delta$ & Value & $\Delta$ & Value & $\Delta$ & Value & $\Delta$ \\ \midrule
\textbf{Baseline} & 2.18 &  & 3.56 &  & 0.82 &  & 1.97 &  & 4.96 &  & 6.56 &  \\ \midrule
\textbf{AWS model}      & \cellcolor{gray!20}1.79 & \cellcolor{gray!20}\textcolor{Green}{-18\%} & \cellcolor{gray!20}3.1 & \cellcolor{gray!20}\textcolor{Green}{-13\%} & 0.52 & \textcolor{Green}{-37\%} & 1.63 & \textcolor{Green}{-17\%} & 4.64 & \textcolor{Green}{-6\%} & 6.11 & \textcolor{Green}{-7\%} \\
\textbf{Azure model}    & 1.35 & \textcolor{Green}{-38\%} & 2.59 & \textcolor{Green}{-27\%} & \cellcolor{gray!20}0.46 & \cellcolor{gray!20}\textcolor{Green}{-44\%} & \cellcolor{gray!20}1.55 & \cellcolor{gray!20}\textcolor{Green}{-21\%} & 3.18 & \textcolor{Green}{-36\%} & 4.53 & \textcolor{Green}{-31\%} \\
\textbf{WhisperX model} & 1.33 & \textcolor{Green}{-39\%} & 2.56 & \textcolor{Green}{-28\%} & 0.52 & \textcolor{Green}{-37\%} & 1.62 & \textcolor{Green}{-18\%} & \cellcolor{gray!20}\textbf{2.31} & \cellcolor{gray!20}\textcolor{Green}{\textbf{-53\%}} & \cellcolor{gray!20}\textbf{3.61} & \cellcolor{gray!20}\textcolor{Green}{\textbf{-45\%}} \\ \midrule
\textbf{Ensemble model} & \textbf{1.32} & \textcolor{Green}{\textbf{-39\%}} & \textbf{2.56} & \textcolor{Green}{\textbf{-28\%}} & \textbf{0.37} & \textcolor{Green}{\textbf{-55\%}} & \textbf{1.44} & \textcolor{Green}{\textbf{-27\%}} & 2.68 & \textcolor{Green}{-46\%} & 3.99 & \textcolor{Green}{-39\%} \\ \bottomrule
\end{tabularx}
\end{table*}

As seen previously in the Fisher test set, the ASR-specific models from Azure and WhisperX performed the best in the datasets transcribed with the same ASR. The exception was the AWS-specific model which performed the worst in the AWS dataset across ASR-specific models.

The ensemble model achieved the best overall performance in this dataset as well. In the WhisperX transcripts, it had comparable performance to the WhisperX model, which performed the best. However, in the Azure and AWS transcripts, it performed better than all ASR-specific models.

\section{Discussion}

Our findings demonstrate that fine-tuning is necessary for LLM-based correction of speaker diarization, as demonstrated previously in Wang et al. 2024 \cite{wang-2024}. Zero-shot model performance was poor, leading to worse diarization accuracy than present at baseline. This may be attributable to a lack of specific adaptation to the task at hand and the varied characteristics of each ASR output. Performance may be improved through use of examples in the prompt, few-shot techniques, or other prompt engineering techniques, as demonstrated in \cite{adedeji-2024}. 

Though fine-tuning led to improved performance, this performance was constrained to transcripts obtained from the same ASR as the transcripts that were used for fine-tuning. This demonstrates that transcripts from different ASR tools are sufficiently varied to impact model performance. The nature of this variance is beyond the scope of this manuscript. However, qualitative observation showed that different ASR tools had different types of errors, e.g., one ASR tool would frequently show inaccurate speaker labeling at the end of sentences, whereas another would mislabel small phrases in the middle of longer speech segments.

In examining the results reported in Wang et al. 2024 \cite{wang-2024} (using USM + turn-to-diarize as their ASR), their best performing model (PaLM 2-S finetuned (hyp2ora)) achieves an impressive 75\% improvement in deltaCP, which significantly outperforms our ASR-specific models. However the best performing open-source model (Llama 3 8B finetuned (mixed) v2; named DiarizationLM 8b Fisher v2 in HuggingFace) achieves approximately 49\% improvement in deltaCP, which is more in line with the performance gains seen in our AWS and Azure models when evaluated on their own test sets, that show 46\% and 54\% improvements respectively.

However, when evaluating DiarizationLM 8b Fisher v2 on transcriptions from unseen ASRs (using our test sets transcribed with AWS, Azure, and WhisperX) the model fails to generalize, leading to a deterioration in diarization performance. This decline in accuracy may be attributed to overfitting to source ASR, large differences in ASR transcription or differences in the amount of diarization errors. Notably, the WhisperX model, which shows the least degradation in performance, had the most comparable baseline diarization error (deltaCP = 4.46) to the ASR used in Wang et al. (deltaCP = 5.71), suggesting that diarization similarity plays a key role in the model’s ability to maintain performance.

The performance of the fine-tuned models was influenced by the distribution of speaker diarization errors in the training dataset. When evaluated on an independent dataset, the model's performance could be negatively affected if the quantity of diarization errors differed significantly from the training dataset, even if transcribed by the same ASR. For instance, the AWS model exhibited the smallest performance improvement on the AWS-transcribed PriMock57 dataset compared to other models. This was likely because AWS-transcribed PriMock57 had a substantially higher baseline rate of speaker diarization errors (deltaCP: 2.18, deltaSA: 3.56) than the AWS-transcribed Fisher test set (deltaCP: 0.93, deltaSA: 2.5). Further analysis revealed that the AWS-specific model became more conservative in the number of speaker changes it implemented, which hindered its performance when evaluated on a dataset with higher error rates.

The ensemble model, with combined weights from each of the individual ASR-specific models, demonstrated better overall performance across ASRs. Evidence suggests the ensemble model may have better adaptability to the different types of speaker mislabeling present across ASRs. Hence, it may be taking a more balanced approach in the amount of speaker corrections it should produce. In the case of correcting diarization errors in the Azure and WhisperX Fisher test transcripts, the ensemble model performed better than the ASR-specific models. It may be that different ASR platforms have common error types present in different distributions. A model that is aware of all such types would perform better than an ASR-specific model tuned to identify the more commonly occurring error types in its ASR.

Improved accuracy through the ensemble model was achieved without compromising on inference time. The merging approach did not change the original architecture. Hence, the combined model had the same size as the individual models. We believe an additional advantage of the ensemble model will be that it may generalize better in transcripts from unfamiliar ASRs, as investigated in \ref{app2}. This will be useful in real-world applications, where a system may be constrained to a specific cloud platform or have a preference for specific ASR tools. The model may be able to improve diarization accuracy as a post-processing step regardless of the ASR tool used to produce the transcript.

We also show that the model is useful on an independent dataset. Notably, this was a clinical dataset, which highlights a major application of such a model i.e. clinical transcriptions. Given the efforts to automate clinical note-taking \cite{wang-2022b, mani-2020, blackley-2019}, such a model could improve those tools significantly, particularly if future efforts involve further fine-tuning specific to conversations that consider the use case.

We acknowledge certain limitations, opening up opportunities for future work. First, a more thorough analysis of the diarization errors produced by each ASR would clarify the need for generalized approaches and help us understand better the types of errors each LLM is more probable to correct. In particular, future work should explore how the results of different ASR models are similar or different, and how LLMs use these variations to improve diarization performance. Especially as it becomes easier to quantify these errors, investigating these differences could provide deeper insights into model adaptability and generalization across ASR outputs. Second, the project was limited to transcripts in English. This restricts diarization correction to a single language. Future work could test this model on transcripts in different languages and also include additional languages in the dataset used for fine-tuning. Third, our experimentation with varied prompts was not extensive \cite{sahoo-2024}. Given we were building on work from Wang et al. 2024 \cite{wang-2024}, we used the same prompts utilized in their work. Future work exploring the transferability of the model could include contextual information (e.g., this is a conversation between a doctor and a patient; this is a phone conversation between a salesperson and a customer) \cite{raju-2018}. Finally, though our project focused on such models serving as post-processing steps, integrating multimodal data will allow for more robust diarization systems. Combining acoustic information with semantic insights from the LLM will enhance the system’s ability to label speech more effectively \cite{park-2024}, especially in complex and noisy environments.

\section{Conclusion}

We investigated the use of fine-tuned LLMs to improve speaker diarization in conversational transcripts with existing speaker labels. Our findings demonstrate that LLMs can markedly improve diarization accuracy when specifically trained to do so. However, if such training is specific to one ASR tool or using a single dataset, it may constrain the model’s performance to transcripts only from that ASR tool and with similar diarization error rates. An ensemble model with knowledge across ASR tools allows for a generalizable and ASR-agnostic diarization correction tool. We believe such a model can be a useful post-processing step in applications that depend on transcription and diarization of conversations with multiple speakers. We have made these models publicly available at HuggingFace at \url{https://huggingface.co/bklynhlth} and built a Python function for easier use of the model available at \url{https://github.com/bklynhlth/openwillis}.

\section{Acknowledgements}
Though referenced throughout the manuscript, the authors would like to acknowledge the authors of Wang et al. 2024 \cite{wang-2024}, whose work formed the foundation of this manuscript. Much of the code used during this work, including the TPST algorithm, critical to model fine-tuning and testing, was originally developed by their group and kindly shared with us for replication.

\section{Author contributions}

\textbf{Georgios Efstathiadis}: Conceptualization, Methodology, Software, Validation, Writing - Original Draft, Visualization \textbf{Vijay Yadav}: Data curation, Software, Investigation, Writing - Review \& Editing \textbf{Anzar Abbas}: Writing - Review \& Editing, Supervision, Project administration

\appendix
\section{Evaluation of zero-shot pre-trained LLMs}
\label{app1}

To determine the optimal model for fine-tuning, we conducted an evaluation of various LLMs using a small sample from the test set. A mix of 15 Fisher test transcriptions was randomly selected, consisting of 5 transcriptions from each ASR (AWS, Azure and WhisperX). After preprocessing and segmenting these transcriptions into manageable prompt-completion pairs, we generated 64 pairs. We focused on small open-source models (fewer than 13 billion parameters), to keep the fine-tuning process computationally feasible and to ensure our results could be made accessible.

We evaluated the performance of Llama2 Chat models (7b and 13b sizes) and Mistral Instruct (7b size v0.2). The following prompt formats were used for each model:

For the Llama2 models:

\begin{tcolorbox}[colback=gray!10, colframe=gray!40, title=]
\textbf{Prompt:}\\

\texttt{<s>[INST] <<SYS>>} \\
In the speaker diarization transcript below, some words are potentially misplaced. Please correct those words and move them to the right speaker. Directly show the corrected transcript without explaining what changes were made or why you made those changes: \\
\texttt{<</SYS>>} \\
\\
\texttt{[ASR transcript] [/INST]} \\
\\
Here is the corrected transcript with the words moved to the right speaker:
\end{tcolorbox}

For the Mistral model:

\begin{tcolorbox}[colback=gray!10, colframe=gray!40, title=]
\textbf{Prompt:}\\

\texttt{<s>[INST]} \\
In the speaker diarization transcript below, some words are potentially misplaced. Please correct those words and move them to the right speaker. Directly show the corrected transcript without explaining what changes were made or why you made those changes: \\
\\
\texttt{[ASR transcript] [/INST]} \\
\\
Here is the corrected transcript with the words moved to the right speaker:
\end{tcolorbox}

The results are presented in Table \ref{tab:base}.

\begin{table}[ht]
\centering
\caption{Zero-shot evaluation of smaller versions of Llama2 and Mistral base models with no fine-tuning on correction of diarization errors.}
\renewcommand{\arraystretch}{1.2}
\setlength{\tabcolsep}{4pt}
\tiny
\begin{tabularx}{0.5\textwidth}{l*{3}{C}}
\toprule
\label{tab:base}
\textbf{} & \multicolumn{3}{c}{\textbf{Mixed transcripts}} \\ \midrule
          & \textbf{deltaCP} & \textbf{deltaSA} & \textbf{WER} \\ \midrule
\textbf{Baseline} & 0.76 & 2.59 & 27.81 \\ \midrule
\textbf{Llama2 7b} & 21.13 & 23.29 & - \\
\textbf{Llama2 13b} & 11.79 & 14.52 & - \\
\textbf{Mistral 7b} & 15.08 & 17.75 & - \\ \bottomrule
\end{tabularx}
\end{table}

Despite the Llama2 13b model demonstrating the best performance, we opted to proceed with the Mistral 7b model for fine-tuning. The Mistral 7b model requires less computational power, making it a more cost-effective solution. Although the Llama2 13b model showed superior performance metrics, the marginal gains did not justify the significantly higher resource demands, especially in a production environment where efficiency and scalability are crucial. By choosing Mistral 7b, we aimed to maintain a high standard of performance while optimizing for ease of deployment.

\section{Generalization to unseen ASR}
\label{app2}

To assess the generalization capabilities of our models, we transcribed the test set using a fourth ASR, without fine-tuning an ASR-specific model on its training data. We transcribed the Fisher test set using the Google Cloud Platform (GCP) transcription service. For the transcription we used the GCP speech to text API (speech\_v1p1beta) with the \textit{latest\_long} model enabling speaker diarization capabilities on, setting the maximum speakers to 2 and specifying the language to English. We filtered the testing data by excluding transcriptions with a single speaker (excluded 8) or those with repeated words issue (excluded 0), resulting in 164 transcriptions.

We evaluated all our fine-tuned expert models and the ASR-specific model on the GCP test data, with the results shown in Table \ref{tab:gcp}.

\begin{table}[ht]
\centering
\caption{Evaluation of fine-tuned models on correction of diarization errors in the GCP Fisher test set.}
\label{tab:gcp}
\renewcommand{\arraystretch}{1.2}
\setlength{\tabcolsep}{4pt}
\tiny
\begin{tabularx}{0.5\textwidth}{l*{4}{C}}
\toprule
\textbf{} & \multicolumn{4}{c}{\textbf{GCP transcripts}} \\ \midrule
          & \textbf{deltaCP} & $\Delta$ & \textbf{deltaSA} & $\Delta$ \\ \midrule
\textbf{Baseline}   & 1.83 &  & 2.98 &  \\ \midrule
\textbf{AWS model}  & 1.82 & \textcolor{Green}{-1\%} & 2.98 & \textcolor{Green}{-0\%} \\
\textbf{Azure model} & 1.93 & \textcolor{Red}{+5\%} & 3.09 & \textcolor{Red}{+4\%} \\
\textbf{WhisperX model} & 2.04 & \textcolor{Red}{+11\%} & 3.21 & \textcolor{Red}{+8\%} \\ \midrule
\textbf{Ensemble model} & \textbf{1.72} & \textcolor{Green}{\textbf{-6\%}} & \textbf{2.86} & \textcolor{Green}{\textbf{-4\%}} \\ \bottomrule
\end{tabularx}
\end{table}

The ASR-specific fine-tuned models did not show any improvement in performance. In fact the Azure and WhisperX models deteriorated baseline performance, while the AWS model achieved a marginal 1\% improvement in deltaCP and 0\% improvement in deltaSA. On the other hand, the ensemble model demonstrated a 6\% improvement in deltaCP and a 4\% improvement in deltaSA.

The GCP test set exhibited a WER of 25.88, the worst transcription performance compared to the other ASRs, leading to higher text incoherence. This likely contributed to the marginal improvements in speaker diarization observed in this analysis.

Despite the small improvement shown by the ensemble model, it represented a significant relative improvement compared to the results from the individual experts that comprised the ensemble. This indicates that while no ASR-specific model was effective at correcting speaker diarization in GCP transcribed data, the ensemble model was able to generalize better and improve performance in an unseen transcription service, indicating the potential of our model to be applied across various ASR services.

\section{Ensemble model - Ablation study}
\label{app3}

The ensemble model was constructed by combining all three ASR-specific models (AWS, Azure, and WhisperX). Each individual model was fine-tuned on transcripts generated by a specific ASR, making them specialized for correcting diarization errors inherent to that ASR. The ensemble approach aimed to leverage the strengths of all three models to improve generalization across different ASRs.

To further analyze the contribution of each ASR-specific model to the ensemble's performance, we conducted an ablation study evaluating ensembles composed of only two ASR-specific models at a time. The following combinations were tested:

\begin{itemize}
    \item AWS + Azure
    \item AWS + WhisperX
    \item Azure + WhisperX
\end{itemize}

Each of these two-model ensembles was evaluated on the Fisher test set from all three ASRs to determine their effectiveness compared to the ASR-specific models and the full three-model ensemble. The results are shown in Table \ref{tab:ablation}.

\begin{table*}[ht]
\centering
\caption{Evaluation of ASR-specific and ensemble models on correction of diarization errors in the Fisher test set. Two-ASR ensemble models showed partial improvements but lacked full generalization. The full ensemble model, incorporating all three ASR-specific models, achieved the best overall performance across different ASRs, demonstrating improved robustness in diarization correction.}
\label{tab:ablation}
\renewcommand{\arraystretch}{1.2}
\setlength{\tabcolsep}{3pt}
\tiny
\begin{tabularx}{\textwidth}{l*{12}{C}}
\toprule
\multirow{3}{*}{} & \multicolumn{4}{c}{\textbf{AWS transcripts}} & \multicolumn{4}{c}{\textbf{Azure transcripts}} & \multicolumn{4}{c}{\textbf{WhisperX transcripts}} \\ \cmidrule(lr){2-13}
                  & \multicolumn{2}{c}{deltaCP} & \multicolumn{2}{c}{deltaSA} & \multicolumn{2}{c}{deltaCP} & \multicolumn{2}{c}{deltaSA} & \multicolumn{2}{c}{deltaCP} & \multicolumn{2}{c}{deltaSA} \\ \cmidrule(lr){2-13}
                  & Value & $\Delta$ & Value & $\Delta$ & Value & $\Delta$ & Value & $\Delta$ & Value & $\Delta$ & Value & $\Delta$ \\ \midrule
\textbf{Baseline} & 0.93 &  & 2.5 &  & 1.91 &  & 3.06 &  & 4.46 &  & 5.77 &  \\ \midrule
\textbf{AWS model}      & \cellcolor{gray!20}0.5 & \cellcolor{gray!20}\textcolor{Green}{-46\%} & \cellcolor{gray!20}2.05 & \cellcolor{gray!20}\textcolor{Green}{-18\%} & 1.3 & \textcolor{Green}{-32\%} & 2.41 & \textcolor{Green}{-22\%} & 4.11 & \textcolor{Green}{-8\%} & 5.39 & \textcolor{Green}{-7\%} \\
\textbf{Azure model}    & 1.04 & \textcolor{Red}{+12\%} & 2.6 & \textcolor{Red}{+4\%} & \cellcolor{gray!20}0.87 & \cellcolor{gray!20}\textcolor{Green}{-54\%} & \cellcolor{gray!20}1.92 & \cellcolor{gray!20}\textcolor{Green}{-37\%} & 3.89 & \textcolor{Green}{-13\%} & 5.12 & \textcolor{Green}{-11\%} \\
\textbf{WhisperX model} & 1.71 & \textcolor{Red}{+84\%} & 3.44 & \textcolor{Red}{+38\%} & 1.46 & \textcolor{Green}{-24\%} & 2.57 & \textcolor{Green}{-16\%} & \cellcolor{gray!20}3.37 & \cellcolor{gray!20}\textcolor{Green}{-24\%} & \cellcolor{gray!20}4.61 & \cellcolor{gray!20}\textcolor{Green}{-20\%} \\ \midrule
\textbf{AWS+Azure model} & \cellcolor{gray!20}0.54 & \cellcolor{gray!20}\textcolor{Green}{-42\%} & \cellcolor{gray!20}2.04 & \cellcolor{gray!20}\textcolor{Green}{-18\%} & \cellcolor{gray!20}0.91 & \cellcolor{gray!20}\textcolor{Green}{-52\%} & \cellcolor{gray!20}1.99 & \cellcolor{gray!20}\textcolor{Green}{-35\%} & 3.45 & \textcolor{Green}{-23\%} & 4.64 & \textcolor{Green}{-20\%} \\
\textbf{AWS+WhisperX model} & \cellcolor{gray!20}0.55 & \cellcolor{gray!20}\textcolor{Green}{-41\%} & \cellcolor{gray!20}2.01 & \cellcolor{gray!20}\textcolor{Green}{-20\%} & 0.97 & \textcolor{Green}{-49\%} & 2.05 & \textcolor{Green}{-33\%} & \cellcolor{gray!20}3.01 & \cellcolor{gray!20}\textcolor{Green}{-33\%} & \cellcolor{gray!20}4.17 & \cellcolor{gray!20}\textcolor{Green}{-28\%} \\
\textbf{Azure+WhisperX model} & 1 & \textcolor{Red}{+8\%} & 2.54 & \textcolor{Red}{+2\%} & \cellcolor{gray!20}0.77 & \cellcolor{gray!20}\textcolor{Green}{-60\%} & \cellcolor{gray!20}1.81 & \cellcolor{gray!20}\textcolor{Green}{-41\%} & \cellcolor{gray!20}2.95 & \cellcolor{gray!20}\textcolor{Green}{-34\%} & \cellcolor{gray!20}4.08 & \cellcolor{gray!20}\textcolor{Green}{-29\%} \\ \midrule
\textbf{Ensemble model} & 0.63 & \textcolor{Green}{-32\%} & 2.15 & \textcolor{Green}{-14\%} & 0.82 & \textcolor{Green}{-57\%} & 1.88 & \textcolor{Green}{-39\%} & 3.15 & \textcolor{Green}{-29\%} & 4.31 & \textcolor{Green}{-25\%} \\ \bottomrule
\end{tabularx}
\end{table*}

The two-ASR ensembles performed better in the datasets which were included in their training sets compared to the third left out ASR, in some cases even surpassing the full ensemble model. This further underscores the challenge of achieving broad generalizability. Notably, the Azure+WhisperX ensemble performed the best in both Azure and WhisperX transcripts but failed to generalize to AWS transcripts, where it not only underperformed but also degraded baseline performance. This suggests that AWS transcripts exhibit distinctive characteristics compared to Azure and WhisperX transcripts, which appear to be more similar to each other. These structural differences likely contributed to the observed performance degradations.

Overall, the full ensemble model (AWS + Azure + WhisperX) provided a more balanced approach to diarization correction, reducing overfitting to any single ASR’s error patterns and ensuring better adaptability across different ASR outputs. Interestingly, the AWS+WhisperX ensemble demonstrated comparable or even superior performance in most cases, suggesting that the significant disparity between AWS and WhisperX transcripts in baseline diarization performance (deltaCP: 0.93 vs. 4.46; deltaSA: 2.5 vs. 5.77) may have encouraged complementary error correction.

These findings highlight the importance of incorporating diverse ASR-specific models into the ensemble that can compensate for each other’s weaknesses. While two-ASR ensembles offered performance gains over individual models, they lacked the full generalization benefits observed in the three-model ensemble. Future work could explore alternative fusion methods and incorporating more ASR-specific models to further enhance diarization correction in unseen ASR scenarios

\bibliographystyle{elsarticle-num-names} 
\bibliography{main}

\end{document}